\documentclass[tightenlines,superscriptaddress,eqsecnum,floats,nofootinbib,showpacs]
{revtex4-1}

\usepackage{amssymb}
\usepackage{stmaryrd}
\usepackage{amsmath}
\usepackage{amsfonts}
\usepackage{mathrsfs}
\usepackage{CJK}
\usepackage{amsmath,amssymb,amsfonts}
\usepackage{graphicx}

\def\be{\begin{equation}}
\def\ee{\end{equation}}
\def\ba{\begin{eqnarray}}
\def\ea{\end{eqnarray}}
\def\nn{\nonumber}

\newcommand{\mubar}{{\bar \mu}} 
\newcommand{\nubar}{{\bar \nu}} 
\newcommand{\abs}[1]{{\left|{#1}\right|}} 
\newcommand{\ket}[1]{\vert{#1}\rangle} 
\newcommand{\bra}[1]{\langle{#1}\vert} 

\newcommand{\sgn}{\mathrm{sgn}} 
\newcommand{\Tr}{\mathrm{Tr}} 
\newcommand{\grav}{\mathrm{gr}} 
\newcommand{\sca}{\mathrm{sc}} 
\newcommand{\kin}{\mathrm{kin}} 
\newcommand{\hil}{\mathcal{H}} 

\begin{document}


\title{ Loop quantum cosmology in 2+1 dimension}

\author{Xiangdong Zhang\footnote{scxdzhang@scut.edu.cn}}
\affiliation{Department of Physics, South China University of
Technology, Guangzhou 510641, China}

\begin{abstract}

As a first step to generalize the structure of loop quantum cosmology to the theories with the spacetime dimension other than four, the isotropic model of loop quantum cosmology in 2+1 dimension is studied in this paper.  We find that the classical big bang singularity is again replaced by a quantum bounce in the model. The similarities and differences between the 2+1 dimensional model and the 3+1 dimensional one are also discussed.
\end{abstract}


\maketitle

\section{Introduction}\label{sec:introduction}
Over the last few decades the three dimensional
gravity has received increasing attentions. The
reasons for this are varied. It is well known that 2+1 dimensional vacuum Einstein gravity does not have any local degree of freedom.  This character is quite different from its 3+1 dimensional counterpart, which is featured by a wealth of local dynamics. However, there still exist black holes solutions in $2+1$ anti-de
Sitter spacetime discovered by Banados,
Teitelboim and Zanelli\cite{BTZ}. These solutions possess certain features reminiscent of
the $3+1$ black holes. In addition, it is often useful to consider a
physical system in lower dimensions as
done in quantum field theory and condense matter physics. Thus it is reasonable to extend
this procedure to gravity. In fact, due to the complicated natures of the 3+1 dimensional gravity that a fully desirable 3+1 quantum gravity is still unavailable.  However, since the 2+1 dimensional gravity is relatively simple, one usually employs it to test the validity of quantization procedure, and it is believed that three dimensional gravity will provide some new insights into a better understanding of the
physically relevant four dimensional gravity both at classical and quantum level.

Loop quantum gravity(LQG) is a theoretical framework trying to unify the principal of general relativity(GR) with the quantum mechanics into a single consistent quantum gravity theory\cite{Ro04,Th07,As04,Ma07}. Many aspects of this theory have been investigated these years. In particular, loop quantum cosmology(LQC), which is a symmetric reduction model of LQG, becomes one of  the most thriving directions of LQG recently\cite{LQC5,Boj,Ash-view,AS11,BCM}. One of the most attractive features of LQC is that the cosmological singularity predicted by classical GR is now naturally replaced by a quantum bounce\cite{APS3,ACS}. Although LQC has made a lot of achievements, nowadays all the discussions are restricted to the four dimensional spacetime . Hence it is interesting to investigate the issue of LQC with the spacetime dimension other than four. On the other hand, 2+1 dimensional LQG has already been well established. Historically, 2+1 dimensional LQG was first established by Ashtekar et. al. in 1989\cite{Ashtekar89}. Afterwards many issues of this theory have been investigated, including the length operator in 2+1 dimension\cite{LFR}, Statistical entropy of Euclidean BTZ black hole\cite{FGNP}, and 2+1 dimensional Euclidean quantum gravity\cite{Thiemann98}. Some 2+1 dimensional  gravity models can even help us to resolve those problems related to 3+1 dimensional gravity. For instance, the 2+1 dimensional Euclidean gravity coupled with a massless scalar field can be related to the 3+1 dimensional static spacetime\cite{Ma02}. Moreover, the 2+1 dimensional cosmology is a rather active topic both from classical and quantum perspective\cite{BM06,GCC03,JK13,Carlip}. The purpose of this paper is to make a first step to investigate the issue of the 2+1 dimensional LQC. In addition, the construction of the three dimensional LQC will help us to answer the following interesting questions: Are there any quantities appearing in the evolution equations of LQC being spacetime dimension dependent? Is there  any universal property of LQC which is independent of spacetime dimension? In this paper we shall give the detailed construction of LQC in 2+1 dimension and try to give a tentative answer to  these questions.

This paper is organized as follows: Section \ref{section2} is divided into two subsections. In subsection A, we first review the classical connection formalism of 2+1 dimensional LQG, and then we use it to derive the cosmological Hamiltonian through symmetric reduction. While in subsection B, we start from the action of 2+1 dimensional
Friedman-Robertson-Walker(FRW) flat Universe, and obtain the corresponding connection formalism by Legendre transformation. In section \ref{section3} we construct quantum theory of 2+1 dimensional LQC and obtain the difference equation which represents the evolution of the Universe. The effective Hamiltonian and the effective equations are obtained In section \ref{section4} and \ref{section5} respectively. Conclusions are given in the last section.

\section{classical theory}\label{section2}

\subsection{classical connection dynamics in 2+1 dimension}\label{sec:section1}
To make this paper self-contained, we first review some basic elements of classical 2+1 dimensional gravity. The connection dynamics of 2+1 dimensional Euclidean gravity with a compact $SU(2)$ guage group is obtained in \cite{Thiemann98}, while the Lorentzian theory with a real $SU(2)$ connection can be realized by Wick transform from Euclidean theory\cite{Phd02}. The difference is that the Lorentzian theory has more difficult Hamiltonian constraint, while the Euclidean theory admits  simpler constraints\cite{Phd02}.  The Ashtekar formalism of 2+1 dimensional gravity constitutes a $SU(2)$ connections $A^i_a$ and densitized dyad $E^a_i=\epsilon^{ab}e_{bi}$ defined on an oriented two dimensional manifold $S$, where $a,b=1,2$ is the spatial indices and $i,j=1,2,3$ denotes $SU(2)$ indices. The commutation relation for the canonical conjugate pairs satisfies \ba
\{A^a_i(x),E^j_b(y)\}=\kappa\gamma\delta^j_i\delta¡­^a_b\delta(x,y)
\ea The 2-metric reads $h_{ab}=e_a^ie_{bi}$. Moreover, the densitized dyad $E^a_i$ satisfies $hh^{ab}=E^a_iE^{ai}$, where $h$ is the determinant of the 2-metric $h_{ab}$. The 2+1 dimensional gravity also has three constraints similar with 3+1 dimensional general relativity\cite{Thiemann98,Ma02} \ba
G_i&=&D_aE^a_i\\
V_a&=&\frac{1}{\kappa\gamma}F_{ab}^iE^b_i\\
H_{gr}&=&\frac{\epsilon^{ijk}E^a_iE^b_j}{2\kappa\sqrt{h}}F_{ab}^k-2(\gamma^2+1)\frac{E^a_{[i}E^b_{j]}}{2\kappa\sqrt{h}}K_a^iK_b^j\label{hamiltong}
\ea where $\kappa=8\pi G$. Note that the Hamiltonian constraint (\ref{hamiltong}) is different from that in \cite{Phd02} with a factor $\frac{1}{\sqrt{h}}$. This is because we want to make the constraint density wight one and thus have a chance
to be densely-defined diffeomorphism covariant operators as explained in\cite{Th07,Thiemann98}. Now we consider an isotropic and homogenous $k=0$ Universe. we choose a fiducial
Euclidean metric  $ {}^oq_{ab}$ on the spatial slice of the isotropic observers and introduce a pair of fiducial orthnormal
dyad and co-dyad as $({}^oe^a_i , {}^o\omega^i_a)$ respectively
such that $ {}^oq_{ab}={}^o\omega^i_a{}^o\omega^i_b$.
Then the physical spatial metric is related to the fiducial by $ q_{ab}=a^2 {}^oq_{ab}$, and its line element can be described by the 2+1 dimensional Friedman-Robertson-Walker (FRW) form
\ba
ds^2=-N^2dt^2+a^2(t)\left(dr^2+r^2d\theta^2\right) \label{lineelement}\nn
\ea
where $a$ is the scale factor. The densitized  dyad $E^a_i=pV_0^{-\frac12}\epsilon^{ab}{}^o\omega_{b,i}
$ and spin connection $A_b^j=cV_0^{-\frac12}{}^o\omega^j_b$. By using the classical expression and comological line elements, one can easily yield $p=a$, and $c=\gamma\dot{a}$. These canonical variables satisfy the following commutation relation\ba
\{c,p\}=\frac{\kappa\gamma}{2}
\ea
For our cosmological case, Gaussian and diffeomophism constraints are satisfied automatically. For the gravity part of Hamiltonian constraint, we use the similar strategy in\cite{LQC5}, i.e. $KKEE$ term is proportional to $FEE$ term, hence, the Hamiltonian constraint (\ref{hamiltong}) reduces to \ba
H_{gr}&=&-\frac{1}{2\kappa\gamma^2}\frac{\epsilon^{ijk}E^a_iE^b_j}{\sqrt{h}}F_{ab}^k
\ea Now, similar to the 3+1 dimensional LQC, we consider a massless scalar field $\phi$ coupled to the gravity field. It is quite obvious that the total Hamiltonian reads
\ba
H_{T}&=&-\frac{1}{2\kappa\gamma^2}\frac{\epsilon^{ijk}E^a_iE^b_j}{\sqrt{h}}F_{ab}^k+\frac{p_\phi^2}{2\sqrt{h}}
\ea Here the $p_\phi$ is the conjugate momentum of scalar field $\phi$. The commutation relation between $\phi$ and $p_\phi$ reads $\{\phi,p_\phi\}=1$. In the cosmological model, this Hamiltonian therefore reduces to\ba
H_T&=&-\frac{1}{\kappa\gamma^2}c^2+\frac{p_\phi^2}{2p^2}\label{hamiltonht}
\ea thus the equation of motion for $p$ reads \ba
\dot{p}=\{p,H_T\}=\frac{1}{\gamma}c
\ea By using the Hamiltonian constraint we can easily yield 2+1 dimensional Friedman equation \ba
H^2=\left(\frac{\dot{p}}{p}\right)^2=\frac{1}{\gamma^2 p^2}c^2=\kappa\frac{p_\phi^2}{2p^4}=\kappa\rho
\ea where the $H=\frac{\dot{a}}{a}$ is the Hubble parameter and the matter density in 2+1 dimension by definition is $\rho=\frac{p_\phi^2}{2p^4}$.

\subsection{Classical theory directly from action}

The connection dynamics discribed in the last section can also be derived directly from action of cosmological case. By using the cosmological line element (\ref{lineelement}), the three dimensional scalar curvature reads \ba
R=2\left(\frac{2\ddot{a}}{N^2a}+\frac{\dot{a}^2}{N^2a^2}-\frac{2\dot{N}\dot{a}}{N^3a}\right)
\ea
Thus the action becomes \ba
S&=&\frac{V_0}{8\pi G}\int dtNa^2\left(\frac{2\ddot{a}}{N^2a}+\frac{\dot{a}^2}{N^2a^2}-\frac{2\dot{N}\dot{a}}{N^3a}\right)\nn\\
&=&-\frac{V_0}{8\pi G}\int dt\frac{\dot{a}^2}{N}
\ea Here we use the fact that $\sqrt{-g}=Na^2r$ and the coordinate volume $V_0=\int drd\theta r$. From now on, we will assume $V_0=1$ by a rescaling. In addition, we also fix the lapse function $N=1$. Thus the resulting gravitational Lagrangian density reads \ba
L_{grav}=-\frac{\dot{a}^2}{\kappa}
\ea where $\kappa=8\pi G$. We take the scale factor $a$ as a canonical variable, then its conjugate momentum reads \ba
p_a=\frac{\partial L_{grav}}{\partial \dot{a}}=-2\frac{\dot{a}}{\kappa }
\ea this canonical conjugate pair satisfies the following commutation relation \ba
\{a,p_a\}=1
\ea The Hamiltonian of the gravitational part is \ba
H_{grav}=\dot{a}p_a-L_{grav}=-\frac{\kappa}{4}p_a^2
\ea In order to match the Ashtekar formalism in the last subsection, we introduce a new set of canonical variables $p=a$ and $c=\gamma\dot{a}=-\frac{\kappa\gamma}{2}p_a$. Note that $ \{a,p_a\}=1$, it is easy to see the Poisson bracket between new variables reads \ba
\{c,p\}=\frac{\kappa\gamma}{2}
\ea and the gravitational part of the Hamiltonian with these new variables becomes \ba
H_{grav}=-\frac{1}{\kappa\gamma^2}c^2
\ea which coincides with the gravitational part of Eq.(\ref{hamiltonht})

\section{Quantum theory}\label{section3}

In order to implement the Hamiltonian constraint at quantum level, we need to rewrite the Hamiltonian constraint with a suitable manner. This is essentially because that the classical Hamiltonian constraint involves inverse of the determinate of 2-metric and thus can not be promoted as a well defined operator on the kinematic Hilbert space. In 3+1 dimensional LQG this difficulty can be overcome through the well know classical identity $\frac12\epsilon^{ijk}\frac{\epsilon_{abc}E^b_jE^c_k}{\sqrt{q}}=\frac{1}{\kappa\gamma}\{A^i_a,V\}$, where the $q$ denotes for the determinant of three metric and $V$ is the volume operator. However, this expression does not work in 2+1 dimension, since we have $\frac{1}{\kappa\gamma}\{A^i_a,V\}=\frac{h_{ab}E^b_i}{\sqrt{h}}$ as noted in \cite{Thiemann98}. Fortunately, also as noted in \cite{Thiemann98}, a deformed formula will help us to resolve this problem.  The key point is introducing a quantity called the degenerate vector\ba
E^i=\frac12\epsilon^{ijk}\epsilon_{ab}E^a_jE^b_k
\ea which can be expressed via Thiemann trick as\ba
E^i=\frac{1}{2(\kappa\gamma)^2}\epsilon_{ijk}\epsilon^{ab}\{A^j_a,V\}\{A^k_b,V\}
\ea On the other hand the $F_{ab}^i$ can be expressed by \ba
F_{ab}^i&=&-2\lim_{Ar_\Box\rightarrow 0}\Tr{\left(\frac{h^{\lambda}_{\Box_{jk}}-1}{\lambda^2}{}^0\omega_a^j{}^0\omega_b^k\tau^i\right)}\nn\\
&=&\epsilon^i_{jk}\frac{\sin^2(\lambda c)}{\lambda^2}{}^0\omega_a^j{}^0\omega_b^k
\ea where we consider a square $\Box_{jk}$ in the
i-j plane, every
edge of the square has length $\lambda V_0$
with respect to the fiducial
metric and the $Ar_\Box$ denotes the area of the square. With these two basic ingredients in hand, and taking into account the fact that the original Hamiltonian constraint can be rewritten in a compact form\cite{Thiemann98} \ba
H_\grav=\frac{F_iE^i}{\sqrt{h}}
\ea with $F_i=\frac{1}{2}\epsilon^{ab}F^i_{ab}$, the gravitational part of Hamiltonian constraint becomes \footnote{we adopt the convention \ba\tau_j&=&-\frac i2\sigma_j\nn\\
\Tr{(\tau_i\tau_j)}&=&-\frac12\delta_{ij}\nn\\
\Tr{(\tau_i\tau_j\tau_k)}&=&-\frac14\epsilon_{ijk}
\ea}\ba
H_\grav&=&-\frac{1}{\kappa\gamma^2}\int d^2x\frac{F_iE^i}{\sqrt{h}}=
\frac{2}{\kappa^3\gamma^4\lambda^4}\epsilon^{ij}\epsilon^{kl}\Tr{\left(h^{\lambda}_{\Box_{ij}}h_k\{h_k^{-1},\sqrt{V}\}h_l\{h_l^{-1},\sqrt{V}\}\right)}
\ea where \ba
h^{\lambda}_{\Box_{ij}}=h_ih_jh_i^{-1}h_j^{-1}
\ea denotes the holonomy along a closed loop $\Box_{ij}$.

To quantize the cosmological model, we first need to construct the
quantum kinematics of 2+1 dimensional cosmology by mimicking the loop
quantum cosmology in 3+1 dimension. This quantum kinematics Hilbert spaces constitutes the so-called polymer-like quantization for gravity part while Schrodinger representation is adopted for scalar field. The
kinematical Hilbert space for the geometry part can be defined as
$\mathcal{H}_{\kin}^{\grav}:=L^2(R_{Bohr},d\mu_{H})$, where
$R_{Bohr}$ and $d\mu_{H}$ are respectively the Bohr
compactification of the real line and Haar measure on it
\cite{LQC5}. On the other hand, for convenience we choose
Schrodinger representation for the scalar field \cite{AS11}. Thus
the kinematical Hilbert space for the scalar field part is defined as
in usual quantum mechanics,
$\mathcal{H}_{\kin}^{\sca}:=L^2(R,d\mu)$. Hence the whole Hilbert
space of 2+1 dimensional LQC is a direct product, $\hil_\kin^{BD}=\hil^\grav_\kin\otimes
\hil^\sca_\kin$. Now let $\ket{\mu}$ be the eigenstates of
 $\hat{p}$ in the kinematical Hilbert space $\mathcal{H}_{\kin}^{\grav}$ such that
 \ba
\hat{p}\ket{\mu}=2\pi G\gamma\hbar\mu\ket{\mu}=\frac{\hbar\kappa\gamma}{4}\mu\ket{\mu}. \nn\ea
Then these eigenstates satisfy orthonormal condition
\ba
\bra{\mu_i}{\mu_j}\rangle=\delta_{\mu_i,\mu_j}\ , \ea
where $\delta_{\mu_i,\mu_j}$ is the Kronecker delta function rather than the Dirac distribution. In 2+1 dimensional quantum gravity, the length, area operators are quantized just like their counterparts in 3+1 dimension, the spectrum of length operator reads\cite{LFR} \ba
L=\kappa\hbar\sum_j\sqrt{j(j+1)}
\ea where the $j$ is half integers. This equation tells us the existence of minimal length gap, which is given by \ba
L=\frac{\sqrt{3}}{2}k\hbar\equiv 4\sqrt{3}\pi\ell_P
\ea Note that the quantization of area refers to
physical geometries in 3+1 dimensional LQC\cite{APS3}, thus we adopt the similar argument in our 2+1 dimensional LQC. we should shrink the loop $\Box_{ij}$ till
the edge of the loop, which is measured by the physical metric
$q_{ab}$, reaches the value of minimal length $L$. Since the physical length of
the elementary cell is $\abs{p}$ and each side of $\Box_{ij}$ is $\lambda$
times the edge of the elementary cell, we use a specific function $\nubar(p)$ to denote $\lambda$, and similar to that in \cite{APS3}, we have \ba
\nubar(p)\abs{p}=L\equiv4\sqrt{3}\pi{\ell}_{\textrm{p}}
\ea
For the convenience of studying quantum dynamics, we define new
variables
\ba v:=\frac{\sqrt{3}}{\gamma}sgn(p)\nubar^{-2},\quad b:=\nubar c, \nn\ea
where
$\nubar=\frac{L}{|p|}$ with
$L=4\sqrt{3}\pi{\ell}_{\textrm{p}}$ being a minimum
nonzero eigenvalue of the length operator \cite{Ash-view}. They also
form a pair of conjugate variables as follows
\ba \{b,v\}=\frac{2}{\hbar}\ .\nn
\ea
It turns out that the eigenstates of
 $\hat{v}$ also forms an orthonormal basis of the kinematic Hilbert space of gravity part $\mathcal{H}_{\kin}^{\grav}$.
We denote
$\ket{\phi,v}$ as the generalized orthonormal basis for the whole Hilbert space
$\hil_\kin$. In the following, we often abbreviate $\ket{\phi,v}$ as $\ket{v}$ in the case of without confusion. The action of volume operator on this basis reads \ba
\hat{V}\ket{v}=\frac{\gamma L^2}{\sqrt{3}}\abs{v}\ket{v}
\ea In $(b,v)$ representation, the Hamiltonian constraint can be reformulated as \ba
H&=&-\frac{1}{\kappa\gamma^2}c^2+\frac{p_\phi^2}{2p^2}\nn\\
&=&-\frac{1}{\sqrt{3}\kappa\gamma}b^2\abs{v}+\left(\frac{\sqrt{3}}{\gamma L^2}\right)\frac{p_\phi^2}{2\abs{v}}\label{shcm}
\ea
At the quantum level, we use the commutator to replace the Poisson bracket to get the exact expression of Hamiltonian constraint \ba
\hat{H}_\grav&=&\frac{8}{\kappa^3\gamma^4\nubar^4\hbar^2}\sin^2(\nubar c)\left(\sin(\frac{\nubar c}{2})\sqrt{V}\cos(\frac{\nubar c}{2})-\cos(\frac{\nubar c}{2})\sqrt{V}\sin(\frac{\nubar c}{2})\right)^2\nn\\
&=&\sin(\nubar c) \hat{F}\sin(\nubar c)
\ea where the action of $\hat{F}$ on a quantum state $\Psi(v)$ is defined by \ba
\hat{F}\Psi(v)=- \frac{1}{\sqrt{3}\kappa\gamma}v^2\left(\sqrt{\abs{v-1}}-\sqrt{\abs{v+1}}\right)^2\Psi(v)\equiv F(v)\Psi(v)
\ea  This operator acts on a quantum state $\Psi(v)\in\hil_\kin$, yielding a similar difference equation as in 3+1 dimensional LQC \ba
\hat{H}_\grav\Psi(v)=f_+(v)\Psi(v+4)+f_0(v)\Psi(v)+f_-(v)\Psi(v-4)\label{Hconstraint}
\ea where \ba
f_+(v)&=& -\frac14F(v+2)= \frac{1}{4\sqrt{3}\kappa\gamma}(\abs{v+2})^2\left(\sqrt{\abs{v+1}}-\sqrt{\abs{v+3}}\right)^2\nn\\
f_0(v)&=& \frac14F(v+2)+\frac14F(v-2)\nn\\
f_-(v)&=&-\frac14F(v-2)
\ea Now we turn to the inverse volume operator. As such, we first have the following classical identity \ba
\abs{p}^{-1/2}=\sgn(p)\frac{8}{3\kappa\gamma\nubar}\Tr{\left(\sum_j\tau^jh_j\{h_j^{-1},V^{1/4}\}\right)}
\ea Note that under the replacement $\{,\}\rightarrow\frac{1}{i\hbar}[,],  $ we have \ba
\Tr{\left(\sum_j\tau^jh_j[h_j^{-1},V^{1/4}]\right)}=\frac32\left(\sin(\frac{\nubar c}{2})V^{\frac14}\cos(\frac{\nubar c}{2})-\cos(\frac{\nubar c}{2})V^{\frac14}\sin(\frac{\nubar c}{2})\right)
\ea Since in classical situation we have $V^{-1}=\abs{p}^{-2}$,  it is easy to see that the action of inverse volume operator on a quantum state $\Psi(v)$ reads \ba
\widehat{V^{-1}}\Psi(v)&=&\left(\frac{4}{\kappa\gamma L\hbar}\right)^4\left(\frac{\gamma L^2}{\sqrt{3}}\right)^3v^2\abs{\abs{v+1}^{\frac14}-\abs{v-1}^{\frac14}}^4\Psi(v)\nn\\
&=&\left(\frac{16\sqrt{3}}{\gamma L^2}\right)v^2\abs{\abs{v+1}^{\frac14}-\abs{v-1}^{\frac14}}^4\Psi(v)\nn\\
&:=&B(v)\Psi(v)
\ea It is noteworthy that in the large $v$ region, the eigen value of the inverse volume operator $\widehat{V^{-1}}$ turns out to be \ba
\left(\frac{\sqrt{3}}{\gamma L^2}\right)\frac{1}{\abs{v}}
\ea

Now we come to the singularity resolution issue. In order to deal with this delicate issue, we take the same strategy adopted in \cite{ACS}. Namely, we first need to make some reasonable simplifications for our quantum Hamiltonian constraint such that the whole system is exactly solvable. Then we discuss the issue of singularity resolution within this exactly solvable formalism\cite{ACS}. According to \cite{ACS}, we first make the following replacements:
\ba B(v)\longmapsto \left(\frac{\sqrt{3}}{\gamma L^2}\right)\frac{1}{\abs{v}},\nn
\ea
and
\ba F(v)\longmapsto- \frac{1}{\sqrt{3}\kappa\gamma}\abs{v}. \nn\ea
The first replacement amounts to
assuming $\mathcal {O}(\frac{1}{\abs{v}})\ll 1$, which in turn implies the validity of the second
replacement.

In the corresponding
quantum theory, we denote quantum state
$\Psi(v)\equiv\Psi(v,\phi)$ for short. Then the simplified
Hamiltonian constraint equation reads \ba
\frac{\partial^2\Psi(v)}{\partial\phi^2}=\hat{\Theta}\Psi(v),\label{hs}\ea
where
\ba
\hat{\Theta}\Psi(v)&=&\frac{\kappa}{2}v \sin(b)v \sin(b)\Psi(v)\nn\\
&=&\frac{\kappa}{8}v\left[(v+2)\Psi(v+4)-2v\Psi(v)+(v-2)\Psi(v-4)\right]\nn\\
&\equiv&(\sum^3_{i=1}\hat{\Theta}_i)\Psi(v).\label{hamilton}\ea
This give us a Klein-Gordon type equation for the quantum dynamics of 2+1 dimensional LQC coupled with a massless scalar field. The physical state should satisfy the ``positive frequency" square root \ba
-i\partial_\phi\Psi(v)=\sqrt{\Theta}\Psi(v).
\ea Here it is noteworthy that  there is a superselection, i.e., for each $\epsilon\in[0.4)$ the states $\Psi(v)$ supported on points $v=4n+\epsilon$ are preserved under dynamics. Thus without loss of generality we will fix $\epsilon=0$ and the inner product between two physical states is given by \ba
(\Psi_1,\Psi_2)_{phy}=\frac{1}{\pi}\sum_{v=4n}\frac{1}{\abs{v}}\bar{\Psi}_1(v)\Psi_2(v)
\ea Note that $\ket{0}$ is excluded out of the physical Hilbert space, since it has zero norm. In addition, $(b,v)$ constitutes a canonical conjugate pair, thus the Fourier transforms $\Psi(b)$ has a support on the interval $(0,\pi)$. \ba
\Psi(b):= \sum_{v=4n}e^{\frac{i}{2}vb}\Psi(v)
\ea the inverse transformation reads \ba
\Psi(v)=\frac{1}{\pi}\int_0^\pi db e^{-\frac{i}{2}vb}\Psi(b)
\ea
Now let us set $\chi(v)=\frac{1}{\pi v}\Psi(v)$, then the constraint becomes a second-order differential equation \ba
\partial^2_\phi\chi(b)=2\kappa\left(\sin(b)\partial_b\right)^2\chi(b)\label{chib}
\ea To make this equation more simpler, we introduce a following new variable $x$  \ba
x=\frac{1}{\sqrt{2\kappa}}\ln\left(\tan(\frac{b}{2})\right)
\ea Then the constraint Eq. (\ref{chib}) becomes the Klein-Gordon equation \ba
\partial^2_\phi\chi(b)=\partial_x^2\chi(b):=-\Theta\chi(b)\label{KG}
\ea The physical Hilbert space consists of positive frequency solutions to Eq. (\ref{KG}), which satisfy \ba
-i\partial_\phi\chi(b)=\sqrt{\Theta}\chi(b)
\ea In fact, we can further decompose the solution into left and right moving sectors $\chi(x)=\chi_L(x_+)+\chi_R(x_-)$, where $x_{\pm}=\phi\pm x$. Moreover, we note that there is a symmetry on the solution $\chi(x)$\cite{ACS}, i.e. $\chi(-x)=-\chi(x)$. Thus we can make a further decomposition $\chi(x)=\frac{1}{\sqrt{2}}\left(F(x_+)-F(x_-)\right)$, with $F(x_{\pm})$ being positive/negative frequency solutions to Eq.(\ref{KG}). The physical inner product now becomes \ba
(\chi_1,\chi_2)_{phy}=i\int_{-\infty}^{\infty}\left[(\partial_x\bar{F}_1(x_+))F_2(x_+)-(\partial_x\bar{F}_1(x_-))F_2(x_-)\right]dx
\ea Thus the expectation value of the volume operator reads \ba
\langle \hat{V}\rangle|_\phi&:=&(\chi,\hat{V}|_\phi\chi)_{phy}=\frac{\gamma L^2}{\sqrt{3}}(\chi,\abs{\hat{v}}\chi)_{phy}\nn\\
&=&i\frac{\gamma L^2}{\sqrt{3}}\int_{-\infty}^{\infty}\left[(\partial_x\bar{F}(x_+))(\hat{v}F(x_+))-(\partial_x\bar{F}(x_-))(-\hat{v}F(x_-))\right]dx\nn\\
&=&\frac{\gamma L^2}{\sqrt{6\kappa}}\int_{-\infty}^{\infty}\left[(\partial_x\bar{F}(x_+))\cosh(\sqrt{2\kappa}x)\partial_xF (x_+)+(\partial_x\bar{F}(x_-))\cosh(\sqrt{2\kappa}x)\partial_xF(x_-)\right]dx\nn\\
&=&\frac{2\gamma L^2}{\sqrt{6\kappa}}\int\abs{\frac{\partial F}{\partial x}}^2\cosh(\sqrt{2\kappa}(x-\phi))dx\nn\\
&=&V_+e^{\sqrt{2\kappa}\phi}+V_-e^{-\sqrt{2\kappa}\phi}\label{Vexpectation}
\ea where \ba
V_{\pm}=\frac{2\gamma L^2}{\sqrt{6\kappa}}\int\abs{\frac{\partial F}{\partial x}}^2e^{\mp\sqrt{2\kappa}x} dx
\ea From Eq.(\ref{Vexpectation}), it is easy to see the expectation value of $\hat{V}$ has a nonzero minimum \ba
V_{min}=2\sqrt{V_+V_-}
\ea Thus we can see that all states undergo a big bounce rather than experience a singularity.  Now we turn to another important observable: matter density $\rho=\langle\rho|_{\phi_0}\rangle$. Note that the classical definition of matter density reads $\rho=\frac{p^2_\phi}{2V^2}$. Thus in order to carry out the expectation value of $\rho$, we first need to know the matrix elements of the observable $\hat{p}_\phi$, which are given by \ba
(F_1,\hat{p}_\phi F_2)_{phy}=\hbar\int dx (\partial_x\bar{F}_1(x))\partial_x F_2(x)
\ea   Combining all the ingredients above, we can readily calculate the expectation value of matter denstiy at the moment of $\phi_0$ with a fixed state $\chi(x)=\frac{1}{\sqrt{2}}\left(F(x_+)-F(x_-)\right)$   \ba
\rho&=&\langle\rho|_{\phi_0}\rangle=\frac{(\langle \hat{p}_\phi\rangle)^2}{2(\langle\hat{V}\rangle)^2}\nn\\
&=&\frac{3\kappa\hbar^2}{4\gamma^2 L^4}\frac{\left[\int dx\abs{\partial_xF}^2\right]^2}{\left[\int dx\abs{\partial_xF}^2\cosh(\sqrt{2\kappa}x)\right]^2}\nn\\
&\leq&\frac{1}{\kappa\gamma^2 L^2}=\rho_c
\ea where the fact $\cosh(\sqrt{\kappa}x)\geq1$ has been used in the second line. Interestingly, the upper-bound of the matter density we obtained here coincides with that  we yield from the effective Friedmann equation(see section \ref{section5}).

\section{Effective Hamiltonian}\label{section4}

The effective description of LQC is a delicate and valuable issue
since it may relate the quantum gravity effects to low-energy
physics. The effective Hamiltonian of LQC has been studied from
canonical perspective\cite{Taveras,DMY,YDM,Boj11} as well as path integral
perspective\cite{ACH102,QHM,QDM,QM1,QM2}.

With the help of
the Hamiltonian constraint equation (\ref{hs}), The effective
Hamiltonian can be derived within the timeless path integral formalism.
In the timeless path integral formalism, the transition amplitude is equivalent to the physical inner product \cite{ACH102,QHM}, i.e.,
\begin{align}
A_{tls}(v_f, \phi_f;~v_i,\phi_i)=\langle v_f,
\phi_f|v_i,\phi_i\rangle_{phy}=\lim\limits_{\alpha_o\rightarrow\infty}
\int_{-\alpha_o}^{\alpha_o}d\alpha\langle
v_f,\phi_f|e^{i\alpha\hat{C}}|v_i,\phi_i\rangle, \label{amplitude}
\end{align}
where $\hat{C}=\hat{\Theta}+\hat{p}_{\phi}^2/\hbar^2$. As already has been shown in Refs.\cite{QHM,QDM}, by multiple group averaging and complete basis inserting, we need to calculate
\begin{align}
\langle v_f, \phi_f|e^{i\sum\limits_{n=1}^N{\epsilon\alpha_n}\hat{C}}|v_i,\phi_i\rangle=\sum\limits_{v_{N-1},...v_1}\int d\phi_{N-1}...d\phi_1\prod\limits_{n=1}^N\langle \phi_n|\langle v_n|e^{i\epsilon\alpha_n\hat{C}}|v_{n-1}\rangle|\phi_{n-1}\rangle.
\label{insert basis}
\end{align}
Note that the action of the constraint operator $\widehat{C}$ now has been separated into  matter part and gravitational part, so the exponential on each kinematical space can be calculated separately. To be more precise
for the matter part one gets
\begin{align}
\langle{\phi_n}|e^{i\epsilon\alpha_n\frac{\widehat{p}^2_\phi}{\hbar^2}}|\phi_{n-1}\rangle
=&\int dp_{\phi_n}\langle{\phi_n}|p_{\phi_n}\rangle\langle p_{\phi_n}|e^{i\epsilon\alpha_n\frac{\widehat{p}^2_\phi}{\hbar^2}}|\phi_{n-1}\rangle\nonumber\\
=&\frac{1}{2\pi\hbar}\int dp_{\phi_n}e^{i\epsilon(\frac{p_{\phi_n}}{\hbar}\frac{\phi_n-\varphi_{n-1}}{\epsilon}
+\alpha_n\frac{{p}^2_{\phi_n}}{\hbar^2})}.
\label{material amplitude}
\end{align}
For the gravity part, note that we have the following identity
\ba \int
d\phi_{n}\bra{\phi_n}\bra{v_n}e^{-i\epsilon\alpha_n\hat{\Theta}}\ket{v_{n-1}}\ket{\phi_{n-1}}=
\delta_{v_n,v_{n-1}}-i\epsilon\alpha_n\int
d\phi_{n}\bra{\phi_n}\bra{v_n}\hat{\Theta}\ket{v_{n-1}}\ket{\phi_{n-1}}+\mathcal
{O}(\epsilon^2). \label{GHamilton}\ea
Then, the matrix elements of $\hat{\Theta}$
can be calculated by using Eq.(\ref{hamilton}), which gives us
\ba
&&\int d\phi_n\bra{\phi_n}\bra{v_n}\hat{\Theta}_1\ket{v_{n-1}}\ket{\phi_{n-1}}\nn\\
&=&
\frac{\kappa}{16}v_{n-1}\frac{v_n+v_{n-1}}{2}
(\delta_{v_n,v_{n-1}+4}-2\delta_{v_n,v_{n-1}}+\delta_{v_n,v_{n-1}-4})\nn\\
&=&\frac{1}{2\pi\hbar}\int d\phi_{n}
dp_{\phi_{n}}e^{i\epsilon(\frac{p_{\phi_{n}}}{\hbar}\frac{\phi_n-\phi_{n-1}}{\epsilon})}\frac{\kappa}{16}v_{n-1}
\frac{v_n+v_{n-1}}{2}
(\delta_{v_n,v_{n-1}+4}-2\delta_{v_n,v_{n-1}}+\delta_{v_n,v_{n-1}-4}),\nn
\ea
With the above results and the following formula
\ba
\frac{1}{4}(\delta_{v_n,v_{n-1}+4}-2\delta_{v_n,v_{n-1}}+\delta_{v_n,v_{n-1}-4})=\frac{1}{\pi}\int_0^\pi
db_ne^{-ib_n(v_n-v_{n-1})/2}\sin^2(b_n),\nn
\ea
Eq.(\ref{GHamilton}) can be expressed as
\ba
&&\int
d\phi_{n}\bra{\phi_n}\bra{v_n}e^{-i\epsilon\alpha_n\hat{\Theta}}\ket{v_{n-1}}\ket{\phi_{n-1}}\nn\\
&=&\frac{1}{2\pi\hbar}\int d\phi_{n}dp_{\phi_{n}}e^{i\epsilon(\frac{p_{\phi_{n}}}{\hbar}\frac{\phi_n-\phi_{n-1}}{\epsilon})}\frac{1}{\pi}\int_0^\pi
db_ne^{-ib_n(v_n-v_{n-1})/2}\left[1-i\alpha_n\epsilon\frac{\kappa}{16}v_{n-1}\frac{v_n+v_{n-1}}{2}4\sin^2b_n\right]. \nn\ea
Collecting all the above ingredients, we can write the transition amplitude  as
\ba
&&A_{tls}(v_f, \phi_f;~v_i,\phi_i)\nn\\
&=&\lim\limits_{N\rightarrow\infty}~~~~\lim\limits_{\alpha_\emph{{No}},...,\alpha_\emph{{1o}}\rightarrow\infty}
\left(\epsilon\prod\limits_{n=2}^N\frac{1}{2\alpha_\emph{{no}}}\right)\int_{-\alpha_\emph{{No}}}^{\alpha_\emph{{No}}} d\alpha_N...\int_{-\alpha_\emph{{1o}}}^{\alpha_\emph{{1o}}} d\alpha_1\nonumber\\
&\times&\int_{-\infty}^{\infty}d\phi_{N-1}...d\phi_1\left(\frac{1}{2\pi\hbar}\right)^N\int_{-\infty}^{\infty}
dp_{\phi_{N}}...dp_{\phi_{1}}\sum\limits_{v_{N-1},...,v_1}~\left(\frac{1}{\pi}\right)^N\int^{\pi}_{0}db_N...db_1\nonumber\\
&\times&\prod\limits_{n=1}^{N}\exp{i\epsilon}\left[
\frac{p_{\phi_{n}}}{\hbar}\frac{\phi_n-\phi_{n-1}}{\epsilon}
-\frac{b_n}{2}\frac{v_n-v_{n-1}}{\epsilon}+\alpha_n \left(\frac{p_{\varphi_n}^2}{\hbar^2}-\frac{\kappa}{16}v_{n-1}\frac{v_n+v_{n-1}}{2}4\sin^2b_n\right)\right].\nn
\ea By taking the `continuum limit' we finally get a path integral formulation as
\ba
&&A_{tls}(v_f, \phi_f;~v_i,\phi_i)\nn\\
&=&c\int \mathcal{D}\alpha\int\mathcal{D}\phi\int\mathcal{D}p_{\phi}\int\mathcal{D}v\int\mathcal{D}b ~~\exp\left(
\frac{i}{\hbar}\int_0^1d\tau \left[p_\phi\dot\phi-\frac{\hbar b}{2}\dot{v}+{\hbar}{\alpha}\left(\frac{p_\varphi^2}{\hbar^2}
-\frac{\kappa}{4}v^2\sin^2b\right)\right]\right),\nn
\ea
where $c$ is an overall constant.
Hence, the effective Hamiltonian constraint in our simplified 2+1 dimensional model can be simply written as
\ba
C_{eff}=-\frac{\kappa}{4}v^2\sin^2b+\frac{p_\varphi^2}{\hbar^2}.\nn\ea
It is easy to see from the above expression that the classical Hamiltonian constraint (\ref{shcm}) can be recovered from $C_{eff}$ up to a factor $\frac{1}{|v|}$ in the large scale limit as $\sin b\rightarrow b$. Thus the above quantum model has correct classical limit.
On the other hand, in order to achieve the effective Hamiltonian constraint for the original model in the previous sections, the proper time of isotropic observers should be respected. The factor $\frac{1}{|v|}$ then has to be multiplied to $C_{eff}$. As a result, we obtain
\ba
H_F=-\frac{1}{\sqrt{3}\kappa\gamma}|v|\sin^2b+\frac{\gamma L^2|v|}{\sqrt{3}}\rho,\nn
\ea
where the matter density is defined by
\ba\rho=\frac{p_\phi^2}{2\abs{p}^4}=\frac{3p_\phi^2}{2v^2\gamma^2 L^4}.\label{density}\ea
Note that the above effective Hamiltonian can also be obtained form the classical Hamiltonian (\ref{shcm}) by the heuristic replacement $b\rightarrow
\sin b$. Hence the classical Hamiltonian constraint can be recovered from the effective $H_F$ in the large scale limit.

\section{Effective equation}\label{section5}
By employing the effective Hamiltonian $H_F$ and symplectic structure of 2+1 dimensional loop quantum cosmology, we can easily get equation of motions for
$v$ and $\phi$ respectively as
\ba
\dot{v}&=&\{v,H_F\}=\frac{4}{\sqrt{3}\hbar\gamma\kappa}|v|\sin
(b)\cos(b),\label{vdot}\\
\dot{\phi}&=&\frac{\sqrt{3}
p_\phi}{\gamma L^2|v|}. \label{phidot1} \ea where $p_\phi$ is a constant of motion, and  $\phi$ can be viewed as an emergent time variable. It is easy to
see that Eq.(\ref{vdot}) leads to \ba
H^2=\left(\frac{\dot{v}}{2v}\right)^2=\frac{4}{3\hbar^2\gamma^2\kappa^2}\sin^2
(b)\cos^2(b).\label{hubble}
\ea On the other hand, the effective Hamiltonian constraint $H_F=0$ can be rewritten as
\ba
-\frac{1}{\sqrt{3}\kappa\gamma}|v|\sin^2b+\frac{\gamma L^2|v|}{\sqrt{3}}\rho=0
\ea which gives us \ba
\sin^2b=\frac{\rho}{\rho_c}
\ea where 2+1 dimensional critical matter density is defined by $\rho_c=\frac{1}{\kappa\gamma^2L^2}$. With the help of this equation, the square of the Hubble parameter can be rewritten as \ba
H^2=\kappa\rho\left(1-\frac{\rho}{\rho_c}\right)\label{effectiveH}
\ea This equation has the same form as in 3+1 dimension except some difference on the coefficient. From Eq. (\ref{effectiveH}), it is easy to see that $\dot{v}=0$ at the point of $\rho=\rho_c$, which reminds us of the existence of a quantum bounce
at that point. To see this is really the case, we can calculate the second derivative of $v$ \ba
\ddot{v}&=&\{\dot{v},H_F\}=-\frac{8}{\sqrt{3}\hbar^2\gamma\kappa}\abs{v}\left(1-2\frac{\rho}{\rho_c}\right)
\ea Obviously, $\ddot{v}=\frac{8}{\sqrt{3}\hbar^2\gamma\kappa}\abs{v}\neq0$ at the point of $\rho=\rho_c$, which implies a quantum bounce
occurs. Moreover£¬ combining Eq. (\ref{effectiveH}) with the continuity equation in 2+1 dimension, $\dot{\rho}+2H(\rho+p)=0$, we can obtain 2+1 dimensional Raychauduri equation with loop quantum correction\ba
\frac{\ddot{a}}{a}=\frac{\ddot{v}}{2v}-H^2=\kappa\rho\left(1-\frac{\rho}{\rho_c}\right)-\kappa(\rho+p)\left(1-\frac{2\rho}{\rho_c}\right)
\ea

\begin{table}\label{table1}
\caption{Some differences between 2+1 dimensional LQC and 3+1 dimensional LQC.}
\begin{ruledtabular}
\begin{tabular}{cccccc}
$ $ & 2+1 dimension & 3+1 dimension \\
\hline
Thiemann trick  & $E^i=\frac{1}{2(\kappa\gamma)^2}\epsilon_{ijk}\epsilon^{ab}\{A^j_a,V\}\{A^k_b,V\}$ & $\frac12\epsilon^{ijk}\frac{\epsilon_{abc}E^b_jE^c_k}{\sqrt{q}}=\frac{1}{\kappa\gamma}\{A^i_a,V\}$    \\
\hline
Geometric variable $p$ & $a$ & $a^2$ \\
\hline
Representation& $\nubar=\frac{L}{p}$ & $\mubar=\sqrt{\frac{\Delta}{p}}$ \\
\hline
Poission bracket between basic variables & $\{c,p\}=\frac{\kappa\gamma}{2}$ & $\{c,p\}=\frac{\kappa\gamma}{3}$ \\
\hline
Inverse volume $V^{-1}$ & $\left(\frac{8}{3\kappa\gamma\nubar}\right)^4\Tr{\left(\sum_j\tau^jh_j\{h_j^{-1},V^{1/4}\}\right)^4}$ & $\left(\frac{4}{\kappa\gamma\mubar}\right)^3\Tr{\left(\sum_j\tau^jh_j\{h_j^{-1},V^{1/3}\}\right)^3}$ \\
\hline
Heuristic replacement& $c\rightarrow\frac{\sin(\nubar c)}{\nubar}$ & $c\rightarrow\frac{\sin(\mubar c)}{\mubar}$ \\
\end{tabular}
\end{ruledtabular}
\end{table}

\section{conclusion}\label{section6}

In this paper, we give a detailed construction of the 2+1 dimensional LQC. We start from the classical Hamiltonian constraint, and then by employing the polymer quantization for the geometry sector we obtain the difference equation which represents the evolution of the 2+1 dimensional Universe. In order to have an effective description of 2+1 dimensional LQC, we generalize the path integral formalism of LQC to 2+1 dimension and use it to derive the effective equations of 2+1 dimensional LQC. Our result shows that the classical singularity is again replaced by a quantum bounce. Now let us be back to the questions raised in the introduction. From the effective equation (\ref{effectiveH}) it is easy to see that in 2+1 dimension the bounce structure is just the same as that in 3+1 dimension. This similarity can be traced back to the holonomy correction. However, another important correction, i.e., the so-called inverse volume correction is quite different from that in 3+1 dimension, thus is explicitly spacetime dimension dependent. Moreover, in order to compare 2+1 dimensional LQC with 3+1 dimensional LQC, we list TABLE I to demonstrate the differences between these two theories in a more transparent way.

Our work opens a window to touch the issues of LQC with the spacetime dimension other than four. In the present paper, we focus mainly on the 2+1 dimensional case. On the other hand, nowadays, higher dimensional cosmology becomes more and more popular. Some higher dimensional cosmological models provide a natural explanation of cosmic accelation\cite{Qiang}. Hence it is also very interesting to investigate the issue of the higher dimensional LQC. Recently, LQG has been generalized to arbitrary spacetime dimension by Thiemann et. al\cite{BTTa,BTTb,BTTc,BTTd}, which in turn offers a possibility to explore the issue of higher dimensional LQC.

It is worth noting that recently a novel numerical technique called Chimera\cite{Singh1,Singh2} developed by Singh et. al. is used to confirm the existence of a quantum bounce beyond the effective equation level in 3+1 dimension, especially for states with a wide spread.  It will be interesting to generalize this formalism to 2+1 dimension. We would like to leave this interesting and delicate topic for future study.

Another interesting topic is to link the LQC with LQG. In 3+1 dimensional case, some efforts have been made towards this direction\cite{AC13a,AC13b}. However, due to the extremely complicated nature of quantum dynamics, a fully satisfactory derivation of LQC from LQG is still lacking. Since it is well known that 2+1 dimensional quantum gravity is an exactly solvable model, we hope it can shed some lights on this important issue.

\begin{acknowledgements}
The author would like to thank Prof. Yongge Ma and Dr. Hongbao Zhang for helpful discussions. This work is supported by NSFC with No.11305063  and
the Fundamental Research Funds for the Central University of China
under Grant No.2013ZM107.

\end{acknowledgements}


\end{document}